**Effects of 2.45 GHz radiofrequency upon Leuconostoc mesenteroides Glucose-6-phosphate dehydrogenase enzymatic activity.**

G. Miño-Galaz [1] * ✉, V. Castro-Fernandez [2], J. Martínez-Oyanedel[3], R. Reeves [4], J. Staforelli-Vivanco[5] * ✉, N. Martínez [4],

[1] Departamento de Ciencias Fisicas, Facultad de Ciencias Exactas, Universidad Andres Bello, Republica 498, Santiago, Chile.
[2] Departamento de Biología, Facultad de Ciencias, Universidad de Chile, Santiago, Chile.
[3] Laboratorio de Bioquímica y Biología Molecular, Facultad de Ciencias Biológicas, Universidad de Concepción, Concepción, Chile.
[4] Centro para la Instrumentación Astronómica, departamento de astronomía, Universidad de Concepción, Concepción, Chile.
[5] Departamento de Física, Facultad de Ciencias Físicas y Matemáticas, Universidad de Concepción, Concepción, Chile.

Correspondence should be addressed to, J. Staforelli and G. Miño-Galaz.

✉ e-mail: jstaforelli@udec.cl, german.mino.galaz@gmail.com

Adress: Av. Esteban Iturra S/N, Concepción, Chile
Phone: +56412661339

Conflict of interest: none

Grant information : (Acknowledgements) G. Miño-Galaz and J. Staforelli acknowledge financial support from fund FONDECYT 1171013. J Staforelli acknowledge partial financial support from fund FONDEF IT24i0064. G. Miño-Galaz also thanks partial support from fund DI-17-20/REG - VRID-UNAB. V. Castro-Fernandez acknowledge partial financial support from FONDECYT 1221667. R. Reeves is supported by the ANID BASAL project FB210003. The authors thanks to Dr. J. Cariñe the facilitation of a preamplifiers from CONICYT PAI CONVOCATORIA NACIONAL SUBVENCIÓN A INSTALACIÓN EN LA ACADEMIA CONVOCATORIA AÑO 2019 Folio (77190088) project. All authors have no conflict of interest to declare.

**Abstract**

In this report we evaluate the effect in the enzyme activity of Glucose 6-phosphate Dehydrogenase from Leuconostoc mesenteroides by irradiation with 2.45 GHz radiofrequency at a power output of 0.1 W during a 91 h period. The results show that the RF irradiation preserves the activity of treated samples of this enzyme with respect to a non-treated sample that instead suffer an increased rate of activity loss. Our estimates indicate that the enzyme activation is due to a non-thermal effect. The results are consistent with reports about the effect of 2.45 GHz radiation upon other enzymatic systems.

## Introduction

An extensive body of experimental research has reported the effects of electromagnetic fields in the radiofrequency (RF) range upon proteins [1-20]. The application of RF technology, in this context, ranges from food pasteurization to cancer treatment, covering a wide variety of activation or deactivation effects on biomolecules [21], among many other effects on several biological targets [22-24,44]. Specific examples of modulatory effects of RF upon proteins are the activation of ionic channels and structural changes in cells at 2.45 GHz [13,14,25,26], structural 3D protein alteration at 1 GHz [5,27], increase protein polymerization at 3.7 MHz [10], structural changes in blood proteins at 50 Hz [15], inhibition of ATP synthesis at 10-100 Hz [28,29], among other cases and effects [8]. In this regard, a highly relevant achievement for human health involves the control of tumor growth using alternating current electric fields (A/C-EF) in the 100-200 kHz regime [1,2,12]. This finding has led to effective commercial devices for the palliative treatment of brain glioblastoma, panchreatic adenocarcinoma treatment and tumor treating in neuro-oncology [2,30,31]. Further studies about the possible mechanism of action of tumoral control growth by A/C-EF have revealed that the polymerization of tubulin protein is affected by A/C-EF radiation [10,32,33]. Other reported specific effects are frequency-dependent increases in enzyme activity [7,20,34].

In this report we evaluate the effect in the enzyme activity of Glucose 6-phosphate Dehydrogenase from Leuconostoc mesenteroides (LmG6PDH; EC 1.1.1.388) by irradiation with 2.45 GHz radio-frequency (RF) at low power output, namely 0.1 W. LmG6PDH is a dual enzyme that catalyzes the oxidation of glucose 6-phosphate with NAD+ or NADP+, producing D-glucono-1,5-lactone 6-phosphate and NADH (or NADPH). LmG6PDH is involved in the first step of the pentose phosphate pathway, and it has been extensively studied from kinetics and structure [35-37]. This enzyme has a random kinetic mechanism for NAD+, which implies it has no strict order for substrates binding, nor product release [35]. Structurally, this enzyme is a homodimer, with subunits of 54 kDa. Each monomer has a large α+β domain and a classic dinucleotide binding domain with the structure βαβ characteristic of Rossman fold [37]. Our experimental setup consists in a room temperature time dependent activity decay of LmG6PDH. Thus, two twin samples are allowed to decay its activity during a 91 h period. One of the samples is exposed to RF radiation tuned at 2.45 GHz at 0.1W of radiation power all times, while the other sample does not suffer exposition. The results show that the RF irradiation is capable to conserve the activity of treated samples of LmG6PDH with respect to a non-treated LmG6PDH sample, that suffer an increased rate of activity loss. The obtained results are in line with reports about the effect of 2.45 GHz radiation upon other enzymes such as peroxidases [9] and dehydrogenases [38].

## Materials and methods

**Expression and purification of recombinant glucose 6-phosphate dehydrogenase from Leuconostoc mesenteroides.**

LmG6PDH gene present into the modified pET-28b vector was used for recombinant expression of the protein. This vector inserted six histidines and a cleavage site for the TEV protease into the Nterminal of the protein. E. coli BL21(DE3) were transformed with the plasmid and grown in Luria Bertani broth containing 35 ug/mL at 37 °C until the OD600 reached ~0.4. Expression of the recombinant protein was induced with 1 mM of Isopropyl-β-D-thiogalactopyranoside overnight. Cells were harvested by centrifugation, suspended in binding buffer (50 mM Tris-HCl pH 7.6, 500 mM NaCl, 20 mM imidazole), and disrupted by sonication. After centrifugation (18514g x 30 min), the soluble fraction was loaded onto a Ni2+-NTA affinity column (HisTrap HP, GE Healthcare, UK). Protein was eluted with a linear gradient between 20 and 500 mM imidazole, and fractions with enzyme activity were pooled. The protein was dialyzed against 50 mM Tris and 1 mM MgCl2, pH 7.5. Precipitation with 3.2 M (NH4)2SO4 was performed and the suspension stored at 4 °C. This leaves the stock solution at 2900 U/ml. The stock of 2900 U/mL of LmG6PDH precipitated with ammonium sulfate corresponds to 4.1 mg/mL in protein concentration with a specific activity of approximately 700 U/mg. The radiofrequency exposure it is done upon dilution of the purified enzyme LmG6PDH recombinantly produced by E. coli. So, an aliquot of the purified enzyme that received the radiofrequency exposure and not the bacterial cells.

**RF irradiation setup.**

The experiments related to the radiofrequency setup (in-bulk) involve a simple synthesized signal amplified, antenna emitted setup, as seen in Figure 1. The box is coated with aluminum foil to contain the radiation inside the system. The Frequency Synthesizer (or Local Oscillator) device generates a 2.45 GHz signal with an output power of +20 dBm. The signal is passed through a coaxial cable into a power amplifier with 8.5 dB gain, and with a maximum output power of +24.5 dBm. The amplifier is then connected to a logarithmic antenna with a directional gain at 2.45 GHz of 4.5 dB. In ideal conditions, the emitted power of the system should be 29 dBm. Considering the attenuation due the connecting cables and adapters, the delivered output power to the sample is 20 dBm (equivalent of 0.1 W). The irradiated sample was located inside the polystyrene box at 1 cm of the emission antenna in a 4 ml sealed glass filled up to 80% with the solution of the protein. An identical control sample was insulated in a secondary polyester box coated with aluminum foil to avoid RF contamination that may arise from the first box and from the environment. Both samples were kept to laboratory temperature along all the measurements. A power distribution simulation inside the irradiation box was performed using Electronic Desktop 2020 software from Ansys company (Ansys, Canonsburg, PA) based on the construction parameters of the setup. According to Figure 1c the field distribution varies across regular values of intensities with no identification of hot spots. Therefore, the exposure upon the sample is essentially homogeneous. To offer visualization of the setup used in our research, photos of it are presented in Figure 1d. The details of the instruments are: The radiofrequency signal is generated with an Agilent N9310A RF signal generator 9 KHz-3 GHz (Keysight, Santa

Rosa, CA).The amplifier is a ZFL-272VH+ high power amplifier (MiniCircuits, Hialeah, FL). The used antenna it is a Log Periodic Antenna for the range of 900-2600 MHz (WA5VJB, Dallas, TX).

**Enzyme activity measurement**

The storage solution of the LmG6PDH, containing 2900 U/ml of enzyme activity is diluted to 0.464 U/ml and separated into two 4ml flask with a o-ring sealing lid. Then the samples are separated in their respective aluminum covered polystyrene boxes, one to be treated with RF at 2.45GHz and the other one without RF (control). The assay solution contains the substrates $NAD^+$ at 2 mM, Glucose-6-phosphate at 2mM, and pH 7.5 Tris-HCL buffer at 50mM of final concentration. The assay solution in its respective cuvette is allowed to thermalize at 25 °C for about 15 min prior to the assay. Prior to the enzymatic measurements the RF source is turned off and both 4 ml flasks with the samples removed of their respective boxes and allowed to rest at a temperature of 25°C for 7 min. 100 μl of enzyme solution is used for each activity measurement completing 1ml in each cuvette, then LmG6PDH reach 0.0464 U/ml of concentration at the moment of measurement. The times of measurement are 0; 42.7; 51.3; 65.7; 75.1 and 91.0 h. The initial velocity is measured following the formation of NADH at 340 nm by 200 s at 25°C in a thermally controlled JASCO V-650 spectrometer. An extinction coefficient of 6220 $M^{-1} cm^{-1}$ for NADH was used, and the enzymatic unit (U) was defined as μmol*$min^{-1}$. Each measurement is repeated 3 times for both control and irradiated cases. Right before each measurement is made, the temperature of each 4ml flask is measured with a laser thermometer. The laser thermometer features a lens designed to concentrate infrared rays, directing them through the device to a detector called a thermopile. Subsequently, the thermopile transforms the received infrared radiation into an electrical signal, which is then showcased as temperature units. The activities and the temperatures for each case are reported in Table 1.

**Results**

An overview of the results for the effect of RF irradiation as a function of time is presented in Figure 2. In the figure, it is possible to observe that in both samples, control and treated, the enzyme activity, or initial velocity, diminishes as function of time. This is an expected result because both samples are kept at room temperature, ~23 °C, during 91 h and a decay by means of denaturation or other forms of inactivation are operative. Initially, both samples start at a common point, with an enzyme activity of 33.9 μM NADH/min. The activity is measured up to 91 h with a slope of activity decay higher for the control sample, reaching a value of 5.4 μM NADH/min at 91 h, while the RF treated sample reaches a value of 11.7 μM NADH/min at 91 h of exposure. The difference in the enzyme velocity of both samples, control and IR-treated, is detected at least at 42 h of exposure. Then the two cases remain separated for all rest of the measurements. The results are also summarized in Table 1.

The results are clear evidence of the effect of RF upon the velocity of LmG6PDH that is always higher for the treated case. This case shows a dose dependent effect that is translated to 2.16 times of incremented activity with respect to the non-treated case at 91 h. A possible cause for the velocity increment is the temperature. Although the radiation power is low (0.1W), it is expected that the 2.45 GHz radiation heat the solvent (water) that contains the protein. As can be observed in table 1 a slightly higher average temperature is observed for the treated cases, with a higher temperature difference of ~1.2 °C for specific measurements. In order to estimate the effect of temperature rise by RF irradiation upon the treated group we used the following rationale. At room temperature, mesophilic enzymes like LmG6PDH follow a change in activity with temperature according to the Arrhenius equation, which allows determining the activation energy of enzymes. The reported activation energy for LmG6PDH is 8.36 Kcal/mol [39]. With these data is possible to calculate an activity ratio Velocity-2/Velocity-1 given by a change in temperature T1 to T2. Thus, for LmG6PDH, there is a variation between 25 to 26 °C results in a V2/V1 = 1.048. In contrast, our experimental measurements ratio in increase of range increments from 1.26 to 2.16 which are much higher than the calculated value of 1.048. Thus, to our best estimate, the observed results are not due to a heating effect of the solvent, but possibly to a non-thermal effect upon the protein system. Non-thermal effects have been described in other protein systems [5,40,41].

## Discussion

The results show that 2.45 GHz affect the activity of the enzyme LmG6PDH. A decay in the activity of both samples as function of time is observed. Nevertheless, the control non-irradiated sample decays at a higher rate. In fact, the measurement at 42 h of exposure show significant differences in both samples, with the irradiated case showing higher enzyme activity. For the rest of the measurements the irradiated sample shows an enhanced activity retention with respect to the non-treated case. The obtained results are in line with reports about the effect of 2.45 GHz radiation upon other enzymes such as peroxidases [9] and dehydrogenases [38]. According to our estimates the effect is due to a non-thermal effect. Although the irradiated sample show increased temperature of 1.2 °C in average with respect to the non-treated case, the higher activity shows in all the measurements may not be explained by a thermal effect. Specially, with a difference of 0.3 °C in average, it would activate the treated case in less than 1.048 times. Opposite to this, an increased activity of 2.16 times is observed for the treated case. According to our estimates an increase of 2.16 times in activity is equivalent to the effect of an increase in temperature of 17 °C, taking 24 °C as reference. So, it may be concluded that 2.45 GHz RF irradiation has a positive effect in the retention of the enzyme activity in the treated case. The origin of this effect may be related with an enhancement of the atomic vibrations of the atoms or chemical groups that compose the enzyme. In particular, the effect of oscillatory electric field upon enzymes and proteins that have been studied by molecular dynamic simulations. In those reports, enhanced protein atom centers oscillations are observed [42,43]. Thus, an enhanced vibratory RF triggered regime may be behind the non-thermal effects and the activation of our target enzyme LmG6PDH. This point deserves further research to be elucidated.

## Conclusion

The 91 h of irradiation at 2.45 GHz at 0.1 W inhibit the decay of activity of LmG6PDH with respect to the non-irradiated case. The results are consistent with reports about the effect of 2.45 GHz radiation upon other enzymatic systems. According to our estimates, our observations are originated by a nonthermal effect. Although the irradiated sample show increased temperature of 1.2 °C in average with respect to the non-treated case, the temperature increase does not explain the enhancement of enzymatic activity in the treated case. The effect may be explained by a non-thermal effect in which the RF radiation enhances protein chemical groups oscillation. This enhanced vibratory RF triggered regimes may be behind the observed non-thermal activation of our target enzyme LmG6PDH.

**Figures and Captions**

Figure 1. (a) Diagram of the RF setup. The RF signal is generated with a tuned Frequency Synthesizer, or Local Oscillator, then amplified and emitted with a commercial log-periodic antenna. The RF isolated container consists of a polystyrene box coated on the inside and outside with aluminum foil to maximize the incident radiation and to avoid contamination from the exterior. The distance to the sample is 1 cm. (b) Log Periodic Antenna. (c) Simulation of the power distribution based on the construction parameters of the setup. It can be clearly seen that the field distribution varies across regular values of intensities with no identification of hot spots. (d) Photo of the irradiation setup. The cap of the irradiation box is not shown in the image.

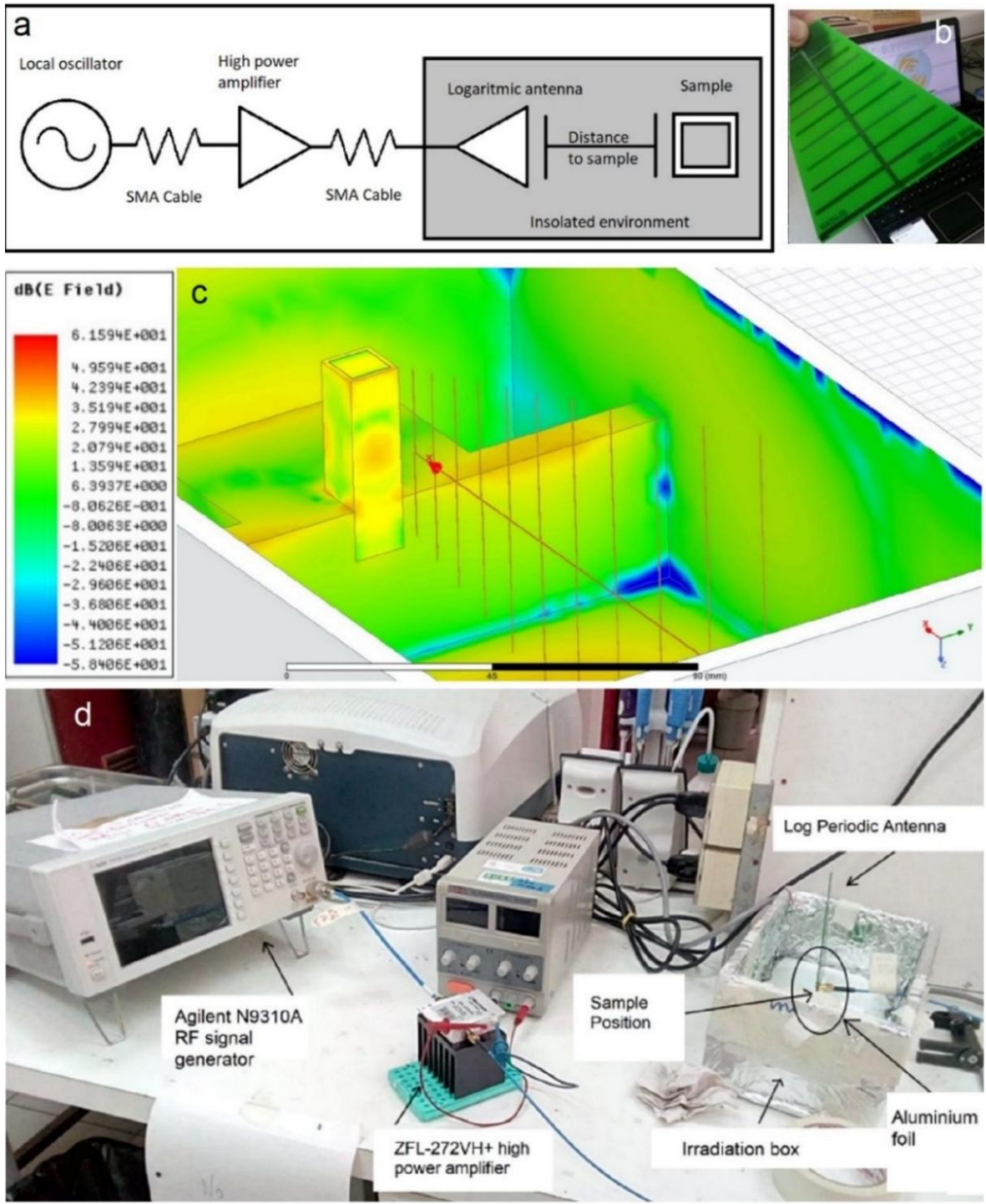

Figure 2.- Glucose 6 phosphate dehydrogenase activity in function of time for control and irradiated cases (each point is represented by the average of 3 independent measurements)

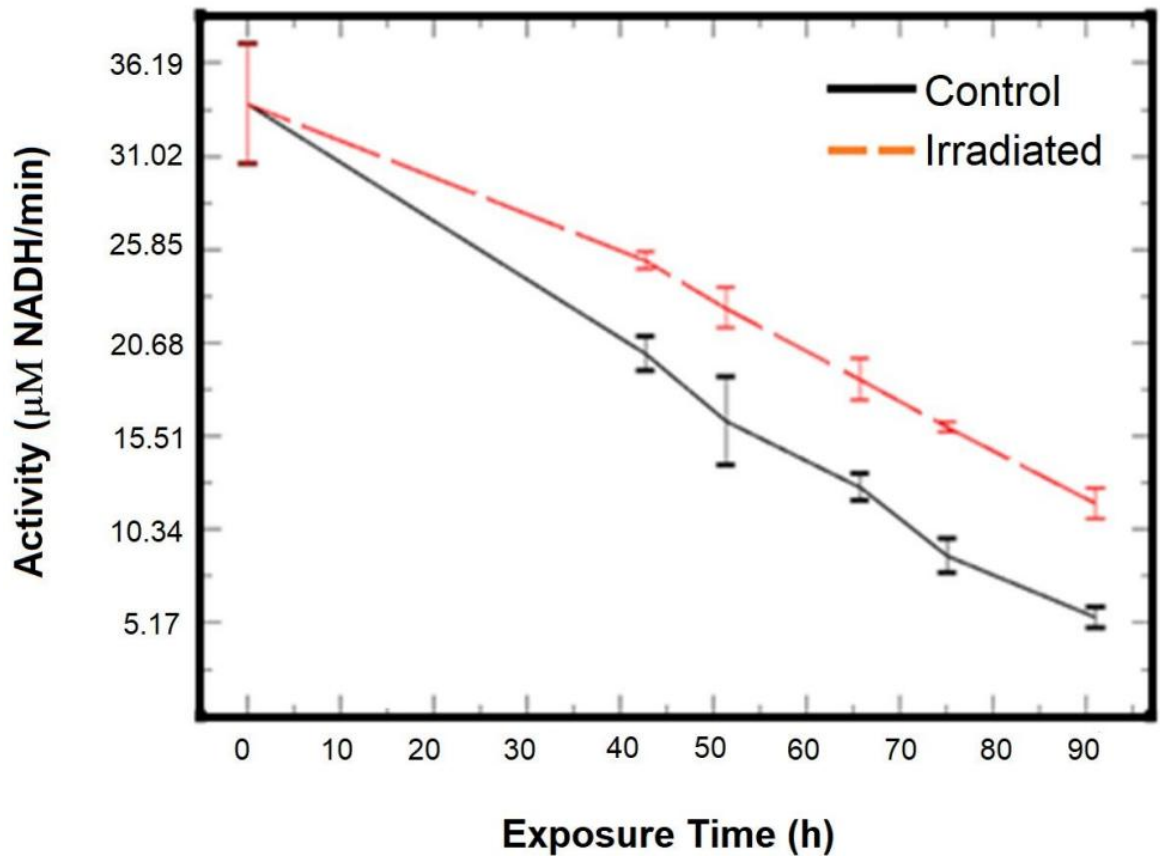

Table 1. Results for the velocity measurement of glucose-6-phosphate dehydrogenase in presence or absence of RF radiation at 2.45 GHz and 0.1 Watt.

| Exposure time[a] | Temp.[b] | Control[c,g] | Temp.[d] | Irradiated[e] | Quotient[f] |
|---|---|---|---|---|---|
| (hrs) | (°C) | (μm NADH/min) (N=3) | (°C) | (μm NADH/min) (N=3) | ([μm NADH/min] Irradiated/ [μm NADH/min] control) |
| 0.0 | 25.0 ± 0.0 | 33.9 ± 3.3 | 25.0 ± 0.0 | 33.9 ± 3.3 | 1.00 |
| 42.7 | 25.0 ± 0.2 | 20.1 ± 1.0 | 25.0 ± 0.0 | 25.2 ± 0.5 | 1.26 |
| 51.3 | 25.5 ± 0.2 | 16.4 ± 2.4 | 26.1 ± 0.3 | 22.6 ± 1.1 | 1.38 |
| 65.7 | 24.4 ± 0.0 | 12.7 ± 0.8 | 24.5 ± 0.1 | 18.7 ± 1.2 | 1.47 |
| 75.1 | 26.3 ± 0.6 | 8.9 ± 0.9 | 26.4 ± 0.2 | 16.0 ± 0.3 | 1.81 |
| 91.0 | 24.6 ± 0.2 | 5.4 ± 0.6 | 24.9 ± 0.1 | 11.7 ± 0.9 | 2.16 |

[a] Exposure time
[b] Temperature of the control sample
[c] Activity of the control Sample
[d] Temperature of the irradiated sample
[e] Activity of the irradiated sample
[f] Quotient of the activities [Irradiated Sample/Control sample]
[g] ± Values represent the standard deviation (SD) for the three independent measurements of temperature or activity

**Figures Captions Page**

Figure 1. (a) Diagram of the RF setup. The RF signal is generated with a tuned Frequency Synthesizer, or Local Oscillator, then amplified and emitted with a commercial log-periodic antenna. The RF isolated container consists of a polystyrene box coated on the inside and outside with aluminum foil to maximize the incident radiation and to avoid contamination from the exterior. The distance to the sample is 1 cm. (b) Log Periodic Antenna. (c) Simulation of the power distribution based on the construction parameters of the setup. It can be clearly seen that the field distribution varies across regular values of intensities with no identification of hot spots. (d) Photo of the irradiation setup. The cap of the irradiation box is not shown in the image.

Figure 2.- Glucose 6 phosphate dehydrogenase activity in function of time for control and irradiated cases (each point is represented by the average of 3 independent measurements)

Table 1. Results for the velocity measurement of glucose-6-phosphate dehydrogenase in presence or absence of RF radiation at 2.45 GHz and 0.1 Watt

## Acknowledgements

G. Miño-Galaz and J. Staforelli acknowledge financial support from fund FONDECYT 1171013. J Staforelli acknowledge partial financial support from fund FONDEF IT24i0064. G. Miño-Galaz also thanks partial support from fund DI-17-20/REG - VRID-UNAB. V. Castro-Fernandez acknowledge partial financial support from FONDECYT 1221667. R. Reeves is supported by the ANID BASAL project FB210003. The authors thanks to Dr. J. Cariñe the facilitation of a preamplifiers from CONICYT PAI CONVOCATORIA NACIONAL SUBVENCIÓN A INSTALACIÓN EN LA ACADEMIA CONVOCATORIA AÑO 2019 Folio (77190088) project. All authors have no conflict of interest to declare.